\documentstyle[NATO,numreferences,epsfig]{CRCKAPB}

\begin{opening}
\title{SEARCH FOR OXYGEN IN COOL DQ WHITE DWARF ATMOSPHERES}
\author{M. Kilic}
\author{D.E. Winget}
\author{T. von Hippel}
\author{D.F. Lester}
\institute{The University of Texas at Austin, Astronomy Department, 1 University Station, C1400, Austin, TX 78712, USA}
\author{D. Saumon}
\institute{Los Alamos National Laboratory, MS-D413, Los Alamos, NM 87544 USA}
\end{opening}
\begin{document}
The existence of carbon in cool He white dwarf (WD) atmospheres has been known for a relatively long time[5]. The presence of carbon in these atmospheres is explained
by convective dredge up of interior carbon [6].
Pelletier {\it et al.} (1986) presented
the first detailed calculations of this process and showed that carbon
diffuses upwards from the core into the base of the He-rich envelope where
it can be dredged up by a surface convection zone.  
As the temperature of the star decreases, more carbon
diffuses upward, and the base of convection zone moves deeper into the
star, further enriching the surface layers with carbon [2].
The diffusion time scales of C and O do not differ by large amounts, and
are found to be essentially the same for some models [3]
. We expect all non-interacting WDs with
$0.45 < M_{wd}/M_\odot < 1.1$
to have C/O cores. Since there is oxygen in
these cores, there is {\it a priori} no reason to prevent the dredge up of
oxygen as well as that of carbon. 
This paper reports new infrared spectroscopic observations of DQ WDs searching for oxygen and describes model atmospheres for these stars.

The WDs chosen for observations were those already known to display the carbon swan bands in their optical spectra and have temperatures between 6000 and 9000K. Our sample is selected from DQs that do not show any abnormalities in their infrared photometry [1]. We have obtained infrared spectroscopy of five WDs at McDonald Observatory in December 2001 using the 2.7m Harlan-Smith Telescope and Rokcam attached to Coolspec.
Throughout the observations, a narrow band CO filter (2.273-2.311 $\mu$m) was utilized in order to decrease the background noise. Figure 1a presents the observed spectra of our sample of WDs.

\begin{minipage}{2.4in}
\epsfig{figure=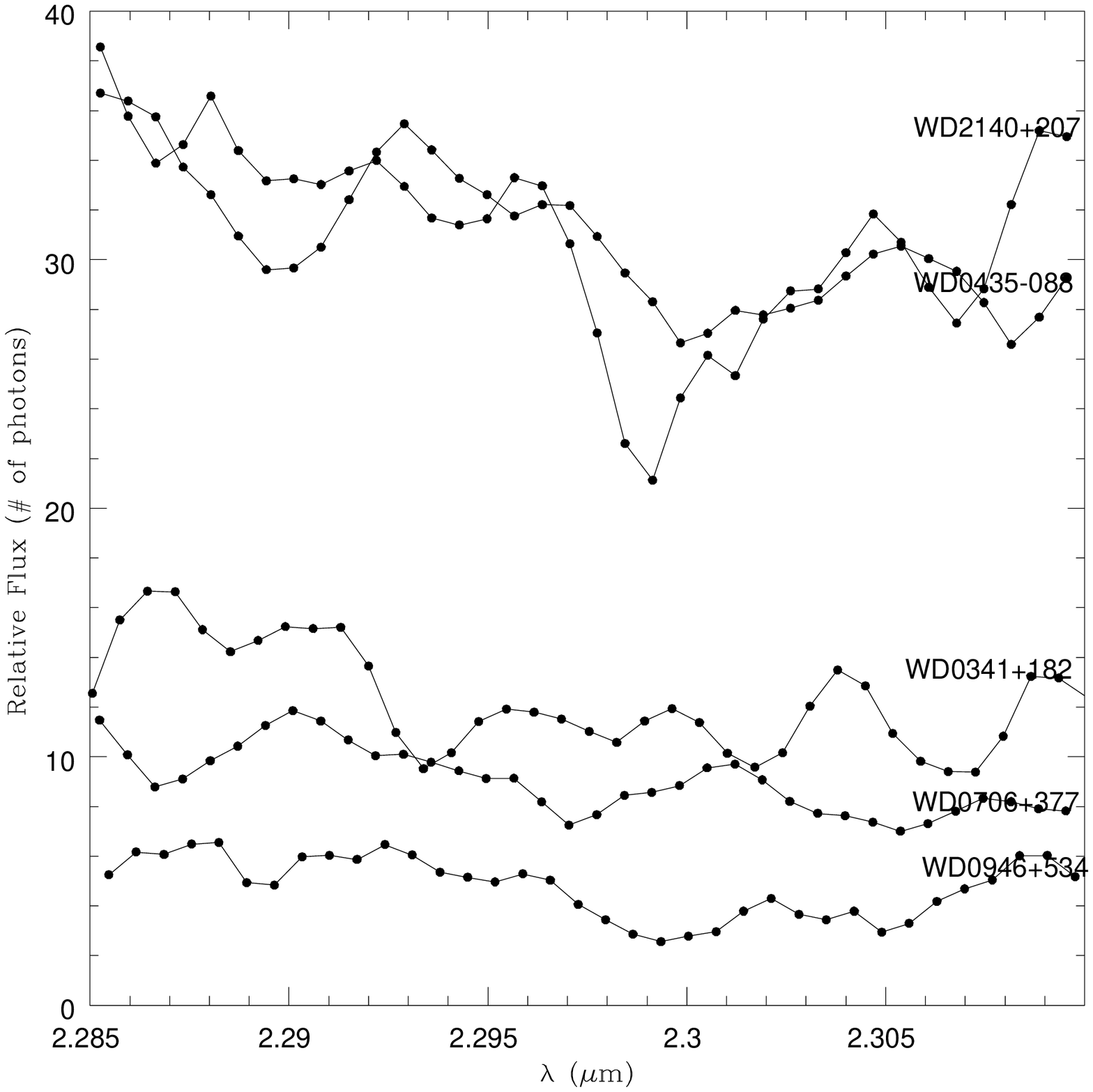,height=2.4truein,angle=0}
\end{minipage}
\begin{minipage}{2.4in}
\epsfig{figure=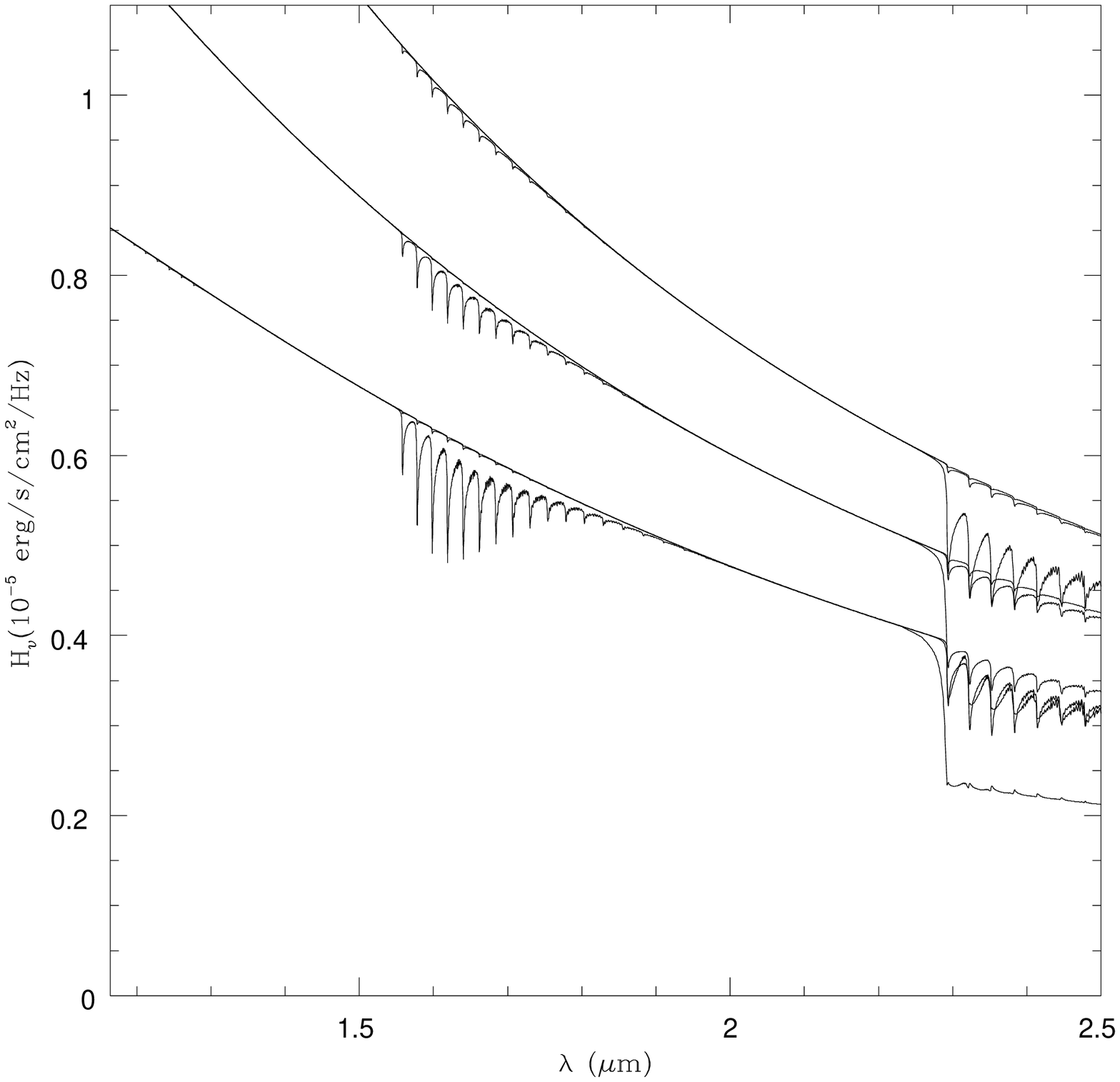,height=2.4truein,angle=0}
\end{minipage}

{\scriptsize {\it Figure 1.} {\it a)} Observed spectra of DQ WDs {\it b)}Effect of CO on pure-He atmosphere models (log g=8, T$_{eff}$=8000, 7000, 6000K (from top to bottom) with three different C/O compositions).}\vspace{0.1in}  

The atmospheric C/O ratio depends in
great part on the C/O ratio at the surface of the WDs' C/O
core.  This remains uncertain in as much as the C$^{12}$($\alpha$,
$\gamma$)$^{16}$O reaction rate is not well known in the astrophysically
important and relevant temperature$-$density domain. 
Models of C and O diffusion and dredge up
in cool He-rich WDs [2] indicate a maximum photospheric abundance
of log $n(O)/n(He)=$ -4.0 to -5.3 for different core compositions.
The most abundant O-bearing molecule is CO, and it has a strong band head at 2.3$\mu$m. Our synthetic spectra calculations for He WDs with different temperatures and different C-O abundances are presented in Fig 1b. The CO abundance is based on a simplified chemical equilibrium calculation that includes CO, C$_2$, C, and O. It appears that CO should be easy to detect for the coolest DQs. This is because at low T$_{eff}$ the ionization of He is decreased, which reduces the background continuum opacity.

Observed spectra of WDs (Fig 1a) show that we might have detected CO in WD2140+207 and WD0435-088. There is also evidence for weak features in the spectra of the other three WDs, but it is impossible to derive the oxygen abundances from these low SNR spectra. Additional follow-up spectroscopy will be obtained in Nov 2002 using the NASA IRTF and Spex.

This work has been supported by NSF grant AST-9876730 and NASA NAG5-9321.

\end{document}